# Improved DWT Based Watermarking Using JPEG-YCbCr*

Mehdi, Khalili
Institute for Informatics and
Automation Problems, National
Academy of Science
Yerevan, Armenia
e-mail: Khalili@ipia.sci.am

David, Asatryan
Institute for Informatics and
Automation Problems, National
Academy of Science
Yerevan, Armenia
e-mail: dasat@ipia.sci.am

## ABSTRACT
In this paper a blind, Secure, imperceptible and robust watermarking algorithm based on wavelet transform domain is proposed in which for more security, the watermark W is converted to a sequence and then a random binary sequence R of size n is adopted to encrypt the watermark, where n is the size of the watermark image. Afterwards, the encrypted watermark sequence W1 is generated by executing exclusive-OR operation on W and R. This generated watermark embeds into low frequency selected coefficients of Y channel wavelet decomposition of JPEG-YCbCr using LSB insertion technique. The experimental results show that the proposed algorithm increases the security and imperceptibility of watermark and has better robustness against wavelet compression and cropping attacks compared to the earlier work in [1].

## 1. INTRODUCTION
With the development of computer technology and network technology, it is more convenient to people to share the information. But at the same time the information security and copyright issues have attracted great attention. Digital watermarking is an effective copyright protection method. It uses embedding technique to embed some important information in carrier to identify the ownership of products [2]. Perceptual invisibility, robustness against different image processing attacks, such as compression and geometric distortions, and ability of watermarking detection without ambiguity are three important requirements of a watermarking scheme [3]. Watermarking techniques can be classified into spatial domain or transform domain techniques [4]. In general, the watermark is often embedded into the transform domain rather than the spatial domain due to the watermark embedded into the transform domain has higher robustness [4]. The obtained results from experiments show the efficiency of presented techniques in the field of digital images watermarking in wavelet transform domain [5] and schemes in low frequency bands will bring about stronger robustness [6]. Also, these results show that watermarking in YCbCr color space versus image processing attacks is more robust than watermarking in RGB color space, and has higher transparency [1]. In this paper, a new blind secure, imperceptible and robust watermarking algorithm is developed by employing a statistic characteristic of low frequency subband coefficients of three levels B-spline wavelet filter. Cause of use of B-spline function wavelet is that, B-spline functions, do not have compact support, and are demonstrated more appropriate orthogonal wavelet base and have better smoothness property for digital image wavelet decomposition [7, 8]. In proposed scheme In order to enhance the security of watermark to the effect of encryption, before watermark embedding into the original data, an encryption algorithm is used for the watermark image processing. The experimental results show that the proposed scheme increases the security and imperceptibility of watermark and has better robustness against wavelet compression and cropping attacks compared to the earlier work in [1].

## 2. JPEG-YCbCr COLOR SPACE
In YCbCr color space, The Y denotes the luminance component. It means that Y shows the brightness (luma). Also both of Cb and Cr represent the chrominance actors. It means that Cb is blue color minus luma (B_Y) and Cr is red color minus luma (R_Y) [9]. The cause of use YCbCr color space was for six reasons [9]. (1) The luminance component (Y) can be adopted to solve the illumination variation problem and it is easy to program because of its independency of the color. (2) According to [10], in YCbCr color space, the skin color cluster is more compact than in other color spaces. (3) YCbCr color space has the smallest overlap between skin and non-skin data in under various illumination conditions. (4) YCbCr color space is very useful in video compression standards (e.g., MPEG and JPEG) [11]. (5) YCbCr color space is a family of color spaces used in video systems. YCbCr color space was defined for standard-definition television use in the ITU-R BT.601 standard for use with digital component video. (6) YCbCr color space accompanied by RGB color space are two primary color spaces used to represent digital component video. JPEG-YcbCr is a rescaling of YCbCr which used in the JPEG image format, with Y, Cb and Cr in [0, 1]. The transform from RGB to JPEG- YCbCr is as follow:

$$\begin{bmatrix} Y \\ Cb \\ Cr \end{bmatrix} = \begin{bmatrix} 0 \\ 0.5 \\ 0.5 \end{bmatrix} + \begin{bmatrix} 0.29890 & 0.58660 & 0.11450 \\ -0.16874 & -0.33126 & 0.50000 \\ 0.50000 & -0.41869 & -0.8131 \end{bmatrix} \begin{bmatrix} R \\ G \\ B \end{bmatrix} \quad (1)$$

While the backward transform from JPEG-YCbCr to RGB is as follow:

$$\begin{bmatrix} R \\ G \\ B \end{bmatrix} = \begin{bmatrix} 1 & 0 & 1.40200 \\ 1 & -0.34414 & -0.71414 \\ 1 & 1.7720 & 0 \end{bmatrix} \times \left( \begin{bmatrix} Y \\ Cb \\ Cr \end{bmatrix} - \begin{bmatrix} 0 \\ 0.5 \\ 0.5 \end{bmatrix} \right) \quad (2)$$

## 3. THE PROPOSED WATERMARKING APPROACH

The current study task of digital watermarking is to make watermarks invisible, secure to human eyes as well as robust to various attacks. Therefore, in the proposed watermarking approach for more security, the watermark W is converted to a sequence and then a random binary sequence R of size n is adopted to encrypt the watermark. By executing exclusive-OR operation on W and R the encrypted watermark sequence W1 is generated and embeded into low frequency selected coefficients of Y channel wavelet decomposition of JPEG-YCbCr using LSB insertion technique. The block diagram of the proposed watermarking approach is shown in figure 2.

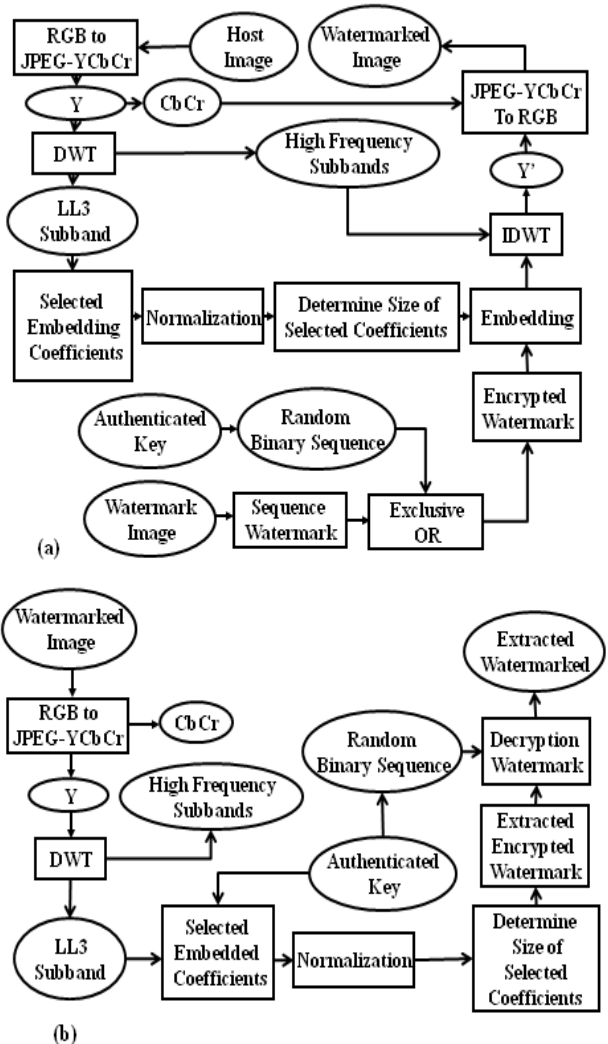

**Figure 2.** Block diagrams of the proposed watermarking approach; (a) Embedding procedure, (b) Extracting procedure.

### 3.1. Watermark Embedding Method

The algorithm for embedding the watermark image in LL3 coefficients of the host image is described as follows:
Step 1: Convert RGB color space of a host image H into JPEG-YCbCr color space using the Eq. 1.
Step 2: Decompose the Y channel into three levels with ten DWT subbands, F(Y). The subband LL3 is taken as the target subband for embedding watermarks.
Step 3: For more security of watermarks, the watermark W is converted to a sequence and then a random binary sequence R of size n is adopted to encrypt the watermark, where n is the size of the watermark image. The encrypted watermark sequence W1 is generated by executing exclusive-OR operation on W and R,

$$W1 = W \oplus R$$

Step 4: Normalize the selected embedding coefficients.
Step 5: Determine size of selected coefficients.
Step 6: Embed watermark W1. For robustness, imperceptibility, and security, the encrypted watermark W1 is embedded in the LSBs of normalized coefficients.
Step 7: Perform Inverse DWT on new Y channel with all changed and unchanged DWT coefficients.
Step 8: Reconvert JPEG-YCbCr color space of the changed host image into RGB color space.
Step 9: Save random binary sequence R and index of the embedded subband as the authenticated key.

### 3.2. Watermark Extraction Method

The embedded watermark can be detected using the stored public key after wavelet decomposition of the watermarked image. The extracting process is described as follows:
Step 1: Convert RGB color space of a watermarked image into JPEG-YCbCr color space using the Eq. 2.
Step 2: Decompose the Y channel into three levels with ten DWT subbands.
Step 3: Re-fetch the authenticated key.
Step 4: Determine size of embedding coefficients.
Step 5: The encrypted watermark image is extracted after normalization of the embedding coefficients.
Step 6: Induce the watermark sequence W' by executing exclusive-OR operation on the sequence W'1 and random binary sequence R,

$$W'1 = W \oplus R$$

## 4. EXPERIMENTAL RESULTS

The proposed perceptual watermarking framework was implemented for evaluating both properties of imperceptibility and robustness. Three 512×512 famous images: Lena, Peppers and Baboon, shown in figure 3(a-c) were taken as the host images to embed a 15×64 binary watermark image, shown in figure 3(d). For the entire test results in this paper, MATLAB R2007a software is used. Also for computing the wavelet transforms, 9-7 biorthogonal spline (Bspline) wavelet filter are used.

The proposed watermarking approach yields satisfactory results in watermark imperceptibility and robustness and improves the earlier work in [1]. The PSNRs of the watermarked images produced by the proposed approach are all greater than 50 dBs, NCs between original watermark images and extracted watermark images are all equal 1, and correlations between host images and watermarked images are all greater than 0.999, which are perceptually imperceptible as shown in figure 4(a-c). Figure 4(d) shows the extracted watermark image from watermarked image. The watermark imperceptibility results are shown in table 1. As it is obvious the proposed scheme improves the Watermark imperceptibility compared to earlier work in [1]. After able to achieve the desired watermark imperceptibility the watermarked images were tested under wavelet compression and cropping attacks to evaluate the robustness property.

### 4.1. Robustness to Wavelet Compression

The process of wavelet image compression is performed as follows: An image is taken, wavelet transform is performed on the digital image, thresholding is done on the digital image, entropy coding is done on the image where necessary, and thus the compression of image is done on the computer. Afterwards, with the compressed image, reconstruction of wavelet transformed image is done, then inverse wavelet transform is performed on the image, thus image is reconstructed.

Here the watermark robustness was examined with wavelet compression. The watermarked images were compressed with different compression thresholds: 3, 5 and 7 using wavelet compression. Then, the watermark was extracted from the compressed watermarked images. The compression results are illustrated in table 2. As it can be seen from the table, for wavelet compression with a threshold 3 the watermark detection is perfect. The recovered watermark for a threshold 5 show a number of detection errors with the maximum error bit rate equal to 3.64% (in Lena) and this only becomes highly noticeable for a threshold 7 with the maximum error bit rate equal to 39.68% (in Lena). Therefore, it can be said that the proposed method yields satisfactory results in robustness against wavelet compression and improves the earlier work in [1].

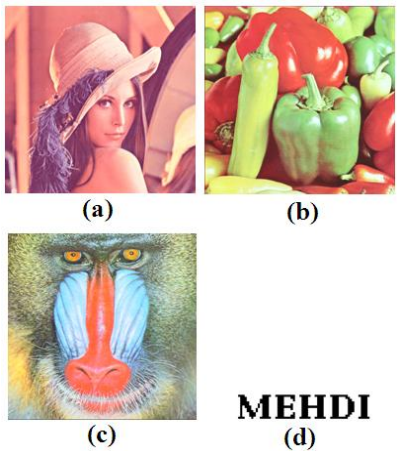

**Figure 3.** The host images for watermarking;
(a-c) Lena, Peppers, and Baboon, (d) Watermark image

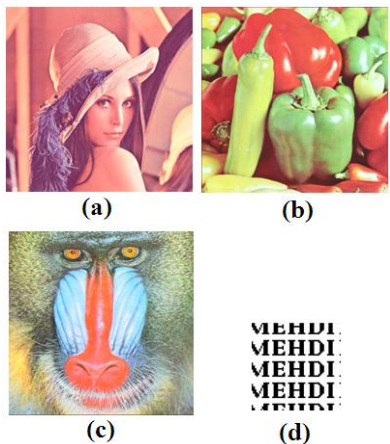

**Figure 4.** (a-c) The watermarked and (d) the extracted watermark images by the proposed approach

### 4.2. Robustness to Cropping Attack

In this examination, two different areas of the watermarked images were cropped and then, the watermark image was extracted from the cropped images. Figure 5 shows the cropped watermarked and extracted watermarks images in different areas and table 3 shows the related obtained results. As it can be seen, the extracted watermarks from cropped watermarked images are recognizable as well with the maximum error bit rate equal 25.57% (in Peppers).

Therefore, the overall robustness of proposed scheme for cropping attack is considered high level, and is higher than earlier work in [5].

**Table 1.** Watermark imperceptibility results

| Image | Lena | Peppers | Baboon |
|---|---|---|---|
| Extracted Watermark | MEHDI | MEHDI | MEHDI |
| PSNR (dB) | 50.49 | 50.50 | 50.48 |
| Correlation | 0.9999 | 0.9999 | 0.9999 |
| NC | 1.00 | 1.00 | 1.00 |
| Error Bit% | 0 | 0 | 0 |

**Table 2.** Obtained results of wavelet compression with different compression thresholds

| Threshold=3 | | | |
|---|---|---|---|
| Image | Lena | Peppers | Baboon |
| Extracted Watermark | MEHDI | MEHDI | MEHDI |
| PSNR (dB) | 48.76 | 48.61 | 49.22 |
| NC | 1.0000 | 1.0000 | 1.0000 |
| Error Bit% | 0.00 | 0.00 | 0.00 |
| Threshold=5 | | | |
| Image | Lena | Peppers | Baboon |
| Extracted Watermark | MEHDI | MEHDI | MEHDI |
| PSNR (dB) | 48.76 | 48.61 | 49.22 |
| NC | 0.9794 | 0.9923 | 0.9965 |
| Error Bit% | 3.64 | 1.35 | 0.62 |
| Threshold=7 | | | |
| Image | Lena | Peppers | Baboon |
| Extracted Watermark | MEHDI | MEHDI | MEHDI |
| PSNR (dB) | 48.76 | 48.61 | 49.22 |
| NC | 0.7835 | 0.8135 | 0.9355 |
| Error Bit% | 39.68 | 34.06 | 11.56 |

| Cropped Area 1 | | | |
|---|---|---|---|
| Image | Lena | Peppers | Baboon |
| Cropped Image | | | |
| Extracted Watermark | | | |
| Cropped Area 2 | | | |
| Image | Lena | Peppers | Baboon |
| Cropped Image | | | |
| Extracted Watermark | | | |

**Figure 5.** The cropped watermarked and extracted watermarked images in different areas

Table 3. Obtained results of cropping attack

| Cropped Area 1 | | | |
|---|---|---|---|
| Image | Lena | Peppers | Baboon |
| NC | 0.9341 | 0.9522 | 0.9713 |
| Error Bit% | 20.16 | 22.41 | 18.93 |
| Cropped Area 2 | | | |
| Image | Lena | Peppers | Baboon |
| NC | 0.9659 | 0.9778 | 0.9841 |
| Error Bit% | 21.81 | 25.57 | 18.66 |

## 5. CONCLUSION

We have proposed a blind watermarking framework for embedding visually recognizable binary watermark in color images which improves the security, watermark imperceptibility and robustness against wavelet compression and cropping attacks compared to the earlier work in [1]. In the proposed approach, a host image was converted into JPEG-YCbCr color space and then, the Y channel was decomposed into wavelet coefficients. Afterwards, the encrypted watermark was embedded into the least significant bit of the low frequency selected coefficients, using LSB insertion technique. The encrypted watermark was generated by executing exclusive-OR operation on W and random binary sequence R; where W is converted watermark to a sequence. The obtained results show that for wavelet compression with a threshold 3 the watermark detection is perfect. The recovered watermark for a threshold 5 show a number of detection errors and this only becomes highly noticeable for a threshold 7. Also, the extracted watermarks from cropped watermarked images are recognizable as well.